\documentclass[twocolumn, aps, prx, reprint, superscriptaddress]{revtex4-2}
\usepackage{staples_style}
\usepackage{colortbl} 
\makeatletter\AtBeginDocument{\let\selectlanguage\@gobble}\makeatother

\begin{document}

\title{Neural-Network Decoders for Qubit Loss}
\author{J. Wilson Staples}
\email{will.staples@princeton.edu}
\affiliation{Princeton Quantum Initiative, Princeton University, Princeton, NJ 08544, USA}
\author{Winston Fu}
\affiliation{Princeton Quantum Initiative, Princeton University, Princeton, NJ 08544, USA}
\author{Yihe Dong}
\affiliation{Princeton Language and Intelligence, Princeton University, Princeton, NJ 08544, USA}
\author{Jeff D. Thompson}
\affiliation{Princeton Quantum Initiative, Princeton University, Princeton, NJ 08544, USA}
\affiliation{Department of Electrical and Computer Engineering, Princeton University, Princeton, NJ 08544, USA}

\date{\today}

\begin{abstract}

Qubit loss is a significant source of error in many quantum hardware platforms, requiring circuits that periodically replace or reset qubits to maintain fault-tolerance. When the reset process includes loss-resolving measurements, this information can improve decoding and substantially reduce overhead compared to the case where this information is not available. Prior work has demonstrated loss-tolerant circuits for the surface code that show significantly improved thresholds and subthreshold scaling by incorporating loss information. However, the computational complexity of loss simulation and decoding, which falls outside of the usual stabilizer framework, has limited extending these studies to high-rate qLDPC codes, which are already challenging to decode in the absence of loss. In this work, we present a loss-aware convolutional neural-network (CNN) decoder framework that circumvents this challenge. Training is enabled by a novel sampler, \texttt{slim}, which achieves a throughput for circuits with Pauli noise and loss that is comparable to the state-of-the-art for pure Pauli noise. On the surface code with loss, the decoder matches the accuracy of computationally expensive most-likely error (MLE) decoders at much lower latency. For the $[[144,12,12]]$ bivariate bicycle (Gross) code, we study a walking syndrome extraction circuit with Pauli and loss errors. For physical error rates $p<1\%$, we observe a more than $10^3$-fold lower logical error rate for a pure loss model, relative to a pure Pauli error model at the same physical error rate, along with significantly enhanced sub-threshold error rate scaling. This demonstrates that loss-dominated error models can also significantly improve performance in a high-rate qLDPC code, and provides a simulation and decoder training platform that can be extended to other codes and algorithmic gadgets.

\end{abstract}

\maketitle

\section{Introduction}
\label{sec:intro}

Fault tolerance is essential to realize large-scale quantum computing. Stabilizer codes use parity checks to detect and correct Pauli errors~\cite{gottesman_class_1996}, with well-understood fault tolerance properties in the presence of stochastic Pauli errors~\cite{shor_fault-tolerant_1996, aharonov_fault-tolerant_1997}. However, physical qubits can also have leakage errors, characterized by qubits leaving the computational subspace. These constitute a significant part of the error budget in many hardware platforms, including superconducting~\cite{miao_overcoming_2023, acharya_quantum_2024}, photonic~\cite{psiquantum_manufacturable_2025, aghaee_rad_scaling_2025}, trapped-ion~\cite{moses_race-track_2023, quinn_high-fidelity_2026, lavasani_quantum_2026}, and neutral-atom~\cite{evered_high-fidelity_2023} qubits. The exact behavior of leakage errors depends on the underlying hardware~\cite{ghosh_understanding_2013, cong_hardware-efficient_2022}. A commonly considered model of this behavior is qubit loss, where leakage errors cause qubits to become computationally inert, effectively skipping subsequent gates~\cite{perrin_quantum_2024, baranes_leveraging_2025, liu_pauli_2026, pavlovich_erasure_2026}. These errors persist and cannot be removed by stabilizer measurements and Pauli corrections alone. Addressing this requires designing circuits with leakage-reduction units, based on teleportation or other primitives to recover or replace qubits~\cite{aliferis_fault-tolerant_2007, fowler_coping_2013, mcewen_relaxing_2023, chow_circuit-based_2024, eickbusch_demonstration_2025}.

At the same time, qubit loss creates an opportunity: if we can detect it independently of the stabilizer parities, we can provide the decoder with additional information to improve logical accuracy~\cite{grassl_codes_1997, sahay_high-threshold_2023, zhang_logical_2026}. We refer to this information as heralds. These can manifest as mid-circuit erasure~\cite{stricker_experimental_2020, ma_high-fidelity_2023, levine_demonstrating_2024, koottandavida_erasure_2024}, providing information about both when and where a loss occurred, or as three-outcome measurements~\cite{chou_superconducting_2024, reichardt_fault-tolerant_2024, scholl_erasure_2023, lis_midcircuit_2023, chow_circuit-based_2024}, which only resolve whether a qubit was lost at some point during its lifetime. In suitable distance-$d$ surface-code circuits, logical failure can require $d$ faults when all faults are heralded losses, compared to $\lceil d/2\rceil$ faults under pure Pauli noise at the same total error rate~\cite{wu_erasure_2022, baranes_leveraging_2025, liu_pauli_2026, perrin_correlated_2026, pavlovich_erasure_2026, bi_foliated_2026, yu_taming_2026, perrin_quantum_2024}.

Existing algorithmic approaches exploit herald information by reweighting the decoding hypergraph for each shot~\cite{suchara_leakage_2015, perrin_quantum_2024, baranes_leveraging_2025, perrin_correlated_2026}. The best known performance requires the use of most-likely error (MLE) solvers~\cite{liu_pauli_2026}, which scale poorly with problem size~\cite{hsieh_np-hardness_2011}. As a consequence, detailed analysis of the performance of quantum error correction with loss has focused on modest-size instances of the surface code, despite the promise of architectures based on higher-rate codes~\cite{bravyi_high-threshold_2024, xu_constant-overhead_2024, webster_pinnacle_2026, bhardwaj_mitten_2026}.

Recent work has demonstrated that machine learning (ML) decoders can achieve accuracy comparable to MLE decoders~\cite{bonilla_ataides_neural_2026}, while also scaling to large code sizes at low latency~\cite{senior_scalable_2025, gu_scalable_2026}. Additionally, ML decoders can incorporate information beyond the syndrome, including analog readout in superconducting surface codes~\cite{varbanov_neural_2025, bausch_learning_2024} and flags for erased fusion outcomes in photonic codes~\cite{barter_machine-learning_2026}. Previous demonstrations of ML decoding with heralded loss have focused on small codes and relatively high logical error rates~\cite{bluvstein_architectural_2025, bonilla_ataides_neural_2026}. A practical obstacle to expanding the scope of these results is the difficulty of simulating large training datasets of error-correcting circuits with qubit loss. 

In this work, we develop convolutional neural network (CNN) decoders for decoding loss noise. This is enabled by a new framework for efficiently sampling from circuits with loss, which performs $1000$ times faster~\cite{luo_ppvm_2026} than existing techniques by extending Pauli-frame sampling to qubit loss. We use this framework to generate training data for our CNN decoders. Our decoders show performance similar to that of computationally expensive MLE decoders on the surface code.

We extend these methods to decode a walking implementation of the $[[144,12,12]]$ bivariate bicycle (Gross) code~\cite{bravyi_high-threshold_2024} under loss noise, for which MLE-based decoding is computationally prohibitive. Using our CNN decoder, we find that biasing noise towards heralded loss significantly improves performance. For physical error rate $p=1\%$, we observe a $2\times10^3$ reduction in logical error rate for a pure loss model, relative to a pure Pauli error model at the same physical error rate. This confirms that loss-dominated error models can also significantly improve performance in a high-rate qLDPC code with loss, and also provides tools to study loss-tolerant circuits and gadgets for high-rate architectures.

\section{Fast Sampling with Loss}
\label{sec:simulation}

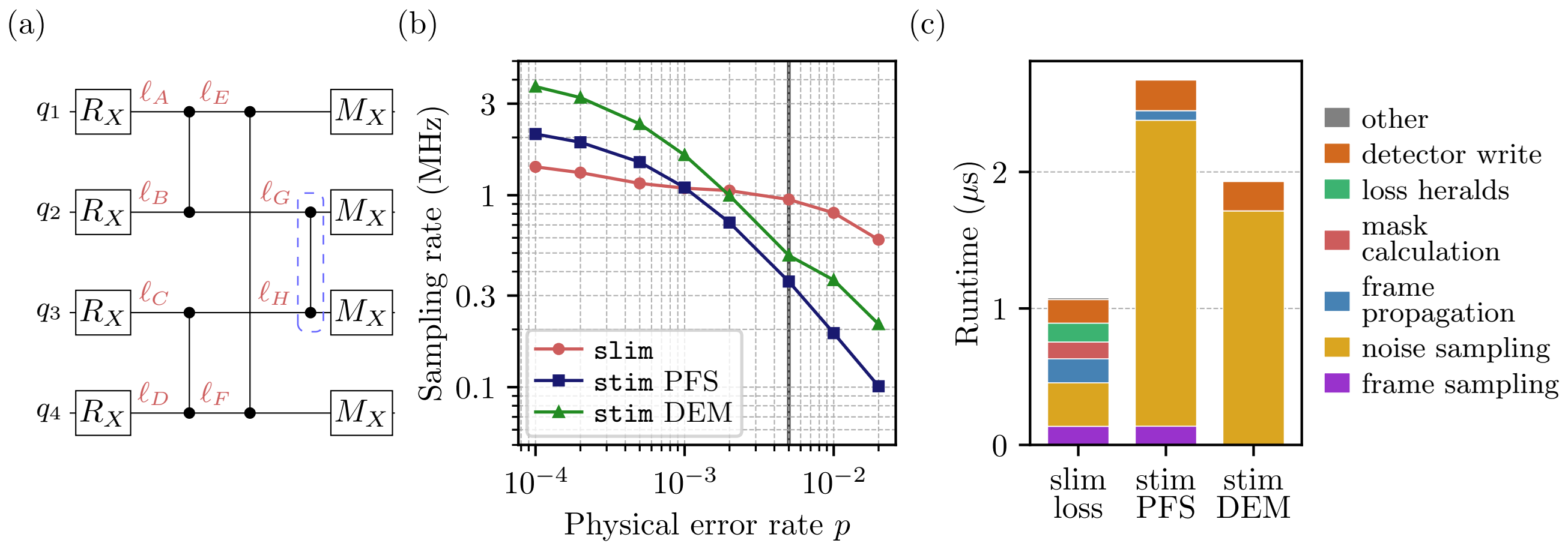
\begin{figure*}[t]
    \centering
    {\setlength{\unitlength}{1bp}%
    \begin{picture}(507.6,176.4)
        \put(0,0){\includegraphics{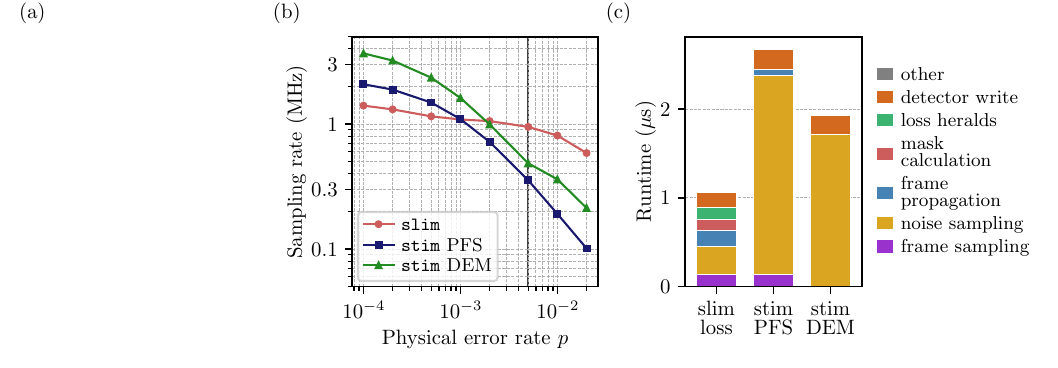}}
        \put(15.23,38.81){
\definecolor{slimred}{HTML}{CD5C5C}
\begin{tikzpicture}[x=1pt, y=1pt,
    wire/.style={line width=0.4pt},
    gate/.style={draw, fill=white, line width=0.4pt, minimum width=16pt, minimum height=14pt,
                 inner sep=1pt},
    dot/.style={circle, fill, inner sep=0pt, minimum size=3.6pt},
    loss/.style={text=slimred, font=\small, inner sep=0pt, anchor=base east},
    group/.style={draw=blue!60, dashed, line width=0.5pt, rounded corners=2.5pt}]
  \def\W{112.4}\def\H{120.4}
  \def\yA{104.5}\def\yB{73}\def\yC{41.5}\def\yD{10}
  \def\xR{21}\def\xI{48}\def\xII{67}\def\xIII{86}\def\xM{102}
  \def\lx{5.5}\def\ly{3.5}
  \useasboundingbox (0,0) rectangle (\W,\H);
  \foreach \y/\q in {\yA/1, \yB/2, \yC/3, \yD/4} {
    \node[anchor=east, font=\small, inner sep=1pt] at (9.5,\y) {$q_{\q}$};
    \draw[wire] (10.5,\y) -- (112.4,\y);
  }
  \foreach \x/\yt/\yb/\lt/\lb in {\xI/\yA/\yB/A/B, \xI/\yC/\yD/C/D,
                                   \xII/\yA/\yD/E/F, \xIII/\yB/\yC/G/H} {
    \draw[wire] (\x,\yt) -- (\x,\yb);
    \node[dot] at (\x,\yt) {};
    \node[dot] at (\x,\yb) {};
    \node[loss] at (\x-\lx,\yt+\ly) {$\ell_{\lt}$};
    \node[loss] at (\x-\lx,\yb+\ly) {$\ell_{\lb}$};
  }
  \draw[group] (\xIII-4,\yC-6) rectangle (\xIII+4,\yB+6);
  \foreach \y in {\yA, \yB, \yC, \yD} {
    \node[gate] at (\xR,\y) {$R_X$};
    \node[gate] at (\xM,\y) {$M_X$};
  }
\end{tikzpicture}%
}
    \end{picture}}
    \caption{Fast sampling of circuits with loss. (a) Loss sites $\ell_A$--$\ell_H$ in a four-qubit circuit. The dashed box marks the final CZ. (b) Throughput against physical error rate $p$ for the $d=11$ moonwalking surface code memory~\cite{pavlovich_erasure_2026} ($12$ rounds) on one CPU thread. Independent loss events before each two-qubit gate are sampled with \texttt{slim}'s masked PFS (red). Independent bit-flips before each entangling gate are sampled with \texttt{stim}'s PFS (navy) and DEM (green). Loss sampling is faster at high $p$ due to using a lookup table for small geometric skips to avoid calculating logarithms. The advantage disappears at low $p$, where runtime becomes dominated by frame sampling and propagation. (c) Runtime budget per shot for $p=5\times10^{-3}$ (vertical line in (b)). These comparisons are done on macOS using an Apple M6 chip.}
    \label{fig:slim}
\end{figure*}

The propagation of Pauli errors through Clifford circuits is deterministic and can be efficiently simulated~\cite{gottesman_heisenberg_1999, aaronson_improved_2004}, enabling high-throughput sampling of circuit-level noise and a compact description of that noise for decoders. This is represented using a Detector Error Model (DEM)~\cite{higgott_sparse_2025}, which assumes that each fault from a set of independent mechanisms $\F$ has a deterministic effect on the syndrome $\sigma \in \FF_2^n$ and that these effects combine linearly. That is, there exists some function $\varphi: \F \to \FF_2^n$ which takes an individual fault to the syndrome that occurs if that fault alone occurs. If a configuration of multiple faults $\{f_1, f_2, \dots, f_m\} \subseteq \F$ occurs, then the resulting syndrome is simply their sum over the binary field--- $\varphi(f_1) + \varphi(f_2)+\cdots +\varphi(f_m)$--- or, equivalently, the XOR of their bit-string representations.

However, Pauli errors can only model physical fault mechanisms that preserve the qubit subspace, excluding leakage errors. The exact behavior of leaked qubits during gates depends on hardware, but a generic and tractable model is qubit loss. We model this as a one-way transition into a non-interacting state $\ket{L}$, effectively turning off subsequent gates~\cite{perrin_quantum_2024, baranes_leveraging_2025, perrin_correlated_2026, liu_pauli_2026, pavlovich_erasure_2026, nishio_qubit_2026, sharma_loss-correcting_2026, luo_ppvm_2026}. 

Loss noise violates both assumptions used for the DEM. First, single loss events have no fixed syndrome. Removing gates alters the way Pauli stabilizers propagate through the circuit, inducing anti-commutation with previously deterministic measurements. Second, loss events do not combine linearly. If two events $\ell_1$ and $\ell_2$ each result in a random outcome for a syndrome element $\sigma_i$, the configuration $\{ \ell_1, \ell_2 \}$ may result in a deterministic outcome for $\sigma_i$. 

Consequently, using a DEM is not an appropriate framework for sampling from circuits with qubit loss. Previous sampling techniques include on-the-fly circuit reconstruction~\cite{perrin_quantum_2024} and per-shot stabilizer tableau simulation~\cite{luo_ppvm_2026} which are $\approx 1000 \times$ slower than sampling from a DEM. We instead use the Pauli Frame Simulator (PFS), which has been used in multiple other published tools for noisy Clifford circuits~\cite{gidney_stim_2021, nvidia_custabilizer_2025, fang_symphase_2023, tuloup_pfsr_2026, hou_quasistim_2026}. This strategy assigns a random ``frame'' bit to the $X$-basis ($Z$-basis) after initialization in $\ket{+}$ ($\ket{0}$). These record how the shot differs from the reference, which can be any measurement outcome that is possible in the absence of noise. 

For generic circuits without loss, this reference can be obtained by simulating the noiseless circuit exactly once with an expensive method such as a stabilizer tableau simulator~\cite{gidney_stim_2021}. However, this strategy is poorly suited for simulating loss noise, where the circuit changes for each shot. Instead, we restrict ourselves to circuits for which the all-zeros measurement is a valid reference, even when any set of entangling gates is skipped. We call this the zero-reference condition (ZRC), and it is satisfied by many error correcting circuits, including those used in this work. How to determine whether a circuit meets the ZRC is discussed in App.~\ref{app:trivial-ref-con}.

For a fixed circuit, we can propagate the sampled frame through the circuit by applying simple Boolean update rules corresponding to each circuit element. For example, if we encounter a $CZ$ gate between qubits $i$ and $j$, then we update the frame as follows:
\begin{equation}
    \label{eq:update-1}
    \begin{aligned}
        z_j &\gets z_j \oplus x_i \\
        z_i &\gets z_i \oplus x_j,
    \end{aligned}
\end{equation}
where $x_a$ and $z_a$ are the respective frame bits for the two tracked Pauli bases on qubit $a$ and $\oplus$ represents the XOR operator. 

For circuits that satisfy the ZRC, we can generalize frame updates to include loss noise by sampling a set of additional random loss indicator bits $\{\ell_i\}$. These denote the occurrence of loss at independent loss locations in the circuit. As we propagate the Pauli frame through the circuit, these indicators can be incorporated into the update rules to faithfully model probabilistic gates. Consider a gate $g$ that updates a Pauli frame according to $y \gets g(y)$ but would be skipped if any of the losses indicated by $\{\ell_1, \ell_2, \dots, \ell_n\}$ occurred. We can define a new update rule:
\begin{equation}
    \label{eq:masked-update}
    y \gets \mu \wedge g(y) \oplus \overline{\mu} \wedge y,
\end{equation}
based on a mask variable:
\begin{equation}
    \mu = \overline{\ell_1}\wedge \overline{\ell_2} \wedge \cdots \wedge \overline{\ell_n},
\end{equation}
where $\wedge$ indicates the Boolean AND operation.

We can use the small example circuit shown in Fig.~\ref{fig:slim}a to illustrate this concretely. This circuit implements a simple error-detection scheme using the parity of measurements on $q_1$ and $q_3$ as well as $q_2$ and $q_4$ as the two syndromes. Here, we consider simulating this circuit under the action of $8$ independent loss events labeled $\ell_A$--$\ell_H$ in the circuit diagram. We focus on the last gate, highlighted with a blue box. If we have a Pauli frame $(x_2, z_2, x_3, z_3)$ on qubits $q_2$ and $q_3$ immediately before the gate, then the gate acts as follows:
\begin{equation}
    \label{eq:ex_gate_transform}
    \begin{gathered}
        \mu = \overline{ \ell_B} \wedge \overline{ \ell_C} \wedge \overline{ \ell_G} \wedge \overline{ \ell_H} \\ 
        \begin{pmatrix} x_2 \\ z_2 \\ x_3 \\ z_3 \end{pmatrix} \gets \begin{pmatrix} x_2 \\ z_2 \oplus x_3 \wedge \mu \\ x_3 \\ z_3 \oplus x_2 \wedge \mu \end{pmatrix}.
    \end{gathered}
\end{equation}
After the gate, we can assign the updated values $z_2$ and $z_3$ to the measurements on $q_2$ and $q_3$, respectively.

This allows us to simulate measurement outcomes in a circuit from random and independently sampled Pauli errors and loss events. Depending on the hardware model, we may want to supply the decoder with loss heralds either in the form of mid-circuit erasures or three-outcome measurements. In either case, the simulation can be easily extended to include this information by including heralds calculated as an OR predicate of indicators. For example, the loss herald $h_2$ representing a three-outcome measurement on qubit $q_2$ can be written as $h_2 = \ell_B \lor \ell_G$.

We adapt \texttt{stim}~\cite{gidney_stim_2021} to implement a stochastic gate PFS as described above in our sampler: \texttt{slim}. Sampling with \texttt{slim} occurs in three broad steps. First, we sample the loss indicators $\{\ell_i\}$ with the geometric skipping technique used in \texttt{stim}, which we specialize to the high loss rates typical of our noise models by storing short skips in a small look-up table instead of computing a logarithm for each one. 

Once we have sampled the loss indicators, the second step is to calculate the mask variables $\{\mu_j\}$ for each entangling gate and the heralds $\{h_k\}$ for each measurement. This can be easily accomplished via stored Boolean relations, which can quickly compute $\{\mu_j\}$ and $\{h_k\}$ for $\{\ell_i\}$. The third and final step is to sample frame values for each qubit initialization, propagate them through the circuit using the masked update rules specified in equation \eqref{eq:masked-update}, and extract measurements from their terminal values. 

We benchmark \texttt{slim} in Fig.~\ref{fig:slim}b using a distance $11$ surface code with a stochastic loss event before each two-qubit gate with probability $p$, and compare throughput to \texttt{stim} using bit-flips with the same probability and circuit locations. We see our loss sampling has throughput similar to bit-flip sampling using a DEM across the range $p \in [10^{-4} , 10^{-2}]$. The advantage in the high $p$ regime is entirely attributable to the look-up table implementation of geometric skipping.

We create a runtime budget for each simulator in Fig.~\ref{fig:slim}c for $p = 5 \times 10^{-3}$. There are three sub-processes in the \texttt{slim} PFS which incur additional runtime compared to the \texttt{stim} PFS: frame forward-propagation (blue) which is less parallel and requires more arithmetic, calculation of element masks (red), and calculation of loss heralds (green). The cumulative effect of these penalties is completely compensated by specializing the noise sampling (yellow), but loss sampling would be strictly slower if the sampling improvement were also used in the Pauli case.

\section{Convolutional Decoder for Loss}
\label{sec:decoder}

\begin{figure*}[t]
    \centering
    {\setlength{\unitlength}{1bp}%
    \begin{picture}(507.84,197.76)
        \put(0,0){\includegraphics{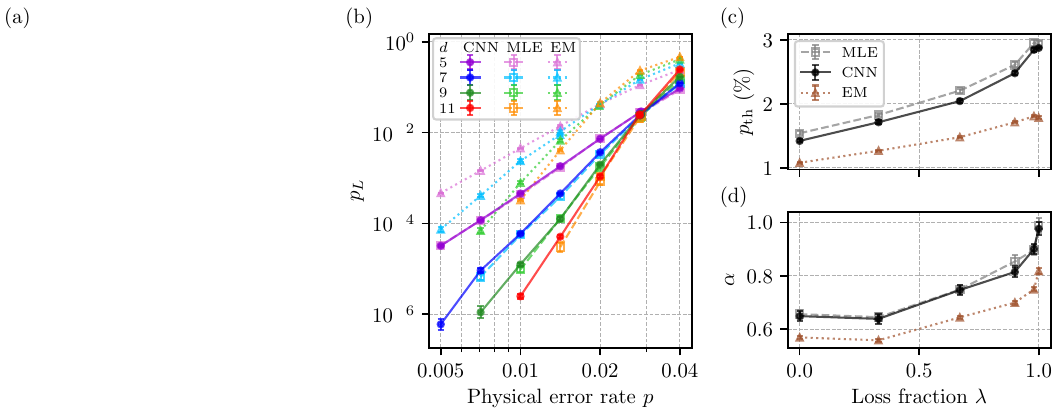}}
        \put(2,5){
\definecolor{archstem}{HTML}{DCE6F5}%
\definecolor{archconv}{HTML}{E4EFDC}%
\definecolor{archread}{HTML}{F6E3D5}%
\definecolor{archhl}{HTML}{CD5C5C}%
\begin{tikzpicture}[x=1pt, y=1pt, >=latex,
    blk/.style={draw, rounded corners=2pt, line width=0.4pt, minimum height=13pt, minimum width=80pt,
                inner sep=1pt, font=\footnotesize, align=center},
    io/.style={font=\footnotesize, inner sep=1.5pt, align=center},
    note/.style={font=\scriptsize, inner sep=1pt, align=left, text=black!70},
    gate/.style={draw, fill=white, line width=0.4pt, minimum width=9pt, minimum height=9pt,
                 inner sep=0.3pt, font=\scriptsize},
    sgate/.style={gate, minimum width=7.4pt, minimum height=7.4pt, font=\tiny},
    dot/.style={circle, fill, inner sep=0pt, minimum size=2.4pt},
    wire/.style={line width=0.4pt},
    cwire/.style={line width=0.4pt, double, double distance=0.6pt},
    flow/.style={->, line width=0.5pt},
    grp/.style={draw=black!45, dashed, line width=0.4pt, rounded corners=3pt}]
  \def\W{150}\def\H{176}
  \useasboundingbox (0,0) rectangle (\W,\H);
  \def\cx{105}       
  \def\ya{155}       
  \def\yk{107}       
  \def\xm{45.5}      
  \def\xj{53}        

  \newcommand{\archmeter}[2]{%
    \node[gate] at (#1,#2) {};
    \draw[line width=0.35pt] (#1-2.8,#2-1.6) arc (150:30:3.23);
    \draw[line width=0.35pt] (#1,#2-2.6) -- (#1+2.2,#2+2.6);}
  \foreach \o/\c in {5/30, 2.5/55} {
    \begin{scope}[shift={(\o,\o)}, black!\c]
      \foreach \y in {14, 7, 0, -7, -14} { \draw[wire] (2,\ya+\y) -- (36,\ya+\y); }
      \node[gate, draw=black!\c] at (40,\ya) {};
    \end{scope}
  }
  \fill[white] (1,\ya-16) rectangle (37,\ya+15.2);
  \foreach \y in {14, 7, -7, -14} { \draw[wire] (2,\ya+\y) -- (36,\ya+\y); }
  \draw[wire, archhl] (2,\ya) -- (36,\ya);
  \foreach \x/\y in {8/14, 15/7, 22/-7, 29/-14} {
    \draw[wire] (\x,\ya) -- (\x,\ya+\y);
    \node[dot] at (\x,\ya) {}; \node[dot] at (\x,\ya+\y) {};
  }
  \archmeter{40}{\ya}
  \draw[wire] (44.5,\ya) -- (\xj,\ya);
  \draw[wire] (48.6,\ya-2.6) -- (50.6,\ya+2.6);

  \begin{scope}[shift={(9,\yk-27-3.4)}]
    \def\s{9}        
    \def\ox{4.2}\def\oy{3.4}   
    \foreach \k in {3,2,1,0} {
      \foreach \i in {0,...,3} {
        \foreach \t in {0,...,5} {
          \pgfmathsetmacro\sh{42-9*\k}\fill[black!\sh] (\i*\s+\k*\ox, \t*\s+\k*\oy) circle (0.85);
        }
      }
    }
    \foreach \k in {2,1,0} {
      \foreach \i in {0,...,2} {
        \foreach \t in {2,...,4} {
          \fill[archhl!55] (\i*\s+\k*\ox, \t*\s+\k*\oy) circle (1.25);
        }
      }
    }
    \fill[archhl] (1*\s+1*\ox, 3*\s+1*\oy) circle (1.7);
    \draw[archhl, line width=0.4pt] (-2.6,2*\s-2.6) rectangle (2*\s+2.6,4*\s+2.6);
    \draw[archhl, line width=0.4pt] (-2.6,4*\s+2.6) -- ++(2*\ox,2*\oy) -- ++(2*\s+5.2,0) -- ++(-2*\ox,-2*\oy);
    \draw[archhl, line width=0.4pt] (2*\s+2.6,2*\s-2.6) -- ++(2*\ox,2*\oy) -- ++(0,2*\s+5.2);
    \draw[->, line width=0.4pt, black!60] (-6,-4) -- (-6,5*\s+2) node[above, note, text=black!60, inner sep=1pt] {$t$};
    \draw[->, line width=0.4pt, black!60] (-4,-6) -- (3*\s+2,-6) node[right, note, text=black!60, inner sep=1pt] {$x$};
    \node[note, anchor=north west, align=center] at (-8,-10) {detectors\\[-1pt]$d\times d\times R$};
  \end{scope}

  \node[blk, fill=archstem] (stem) at (\cx,\ya) {input stem};
  \draw[flow] (\xj,\ya) -- (stem);
  \node[io] (nin) at (\cx,172) {$p,\lambda$};
  \draw[flow] (nin) -- (stem);

  \node[blk, fill=archconv] (down) at (\cx,126) {norm, SiLU, $H\!\to\!H/4$};
  \node[blk, fill=archconv] (conv) at (\cx,\yk) {$3{\times}3{\times}3$ kernel};
  \node[blk, fill=archconv] (up)   at (\cx,88)  {$H/4\!\to\!H$};
  \node[circle, draw, line width=0.4pt, inner sep=0pt, minimum size=8pt, font=\scriptsize] (add) at (\cx,71) {$+$};
  \coordinate (split) at (\cx,140);
  \draw[flow] (stem) -- (down);
  \draw[flow] (down) -- (conv);
  \draw[flow] (conv) -- (up);
  \draw[flow] (up) -- (add);
  \fill (split) circle (0.9);
  \draw[flow] (split) -- (\cx-43.5,140) -- (\cx-43.5,71) -- (add);
  \draw[grp] (\cx-46.5,63) rectangle (\cx+44,144.5);
  \node[font=\footnotesize, anchor=south east, inner sep=2pt] at (\cx+44,63) {$\times L$};
  \draw[archhl, line width=0.4pt, densely dotted] (41.5,\yk) -- (conv.west);

  \node[blk, fill=archread] (pool) at (\cx,48) {$1{\times}2{\times}2$ kernel, mean};
  \node[blk, fill=archread] (head) at (\cx,29) {readout MLP};
  \node[io] (out) at (\cx,9) {logit of logical flip};
  \draw[flow] (add) -- (pool);
  \draw[flow] (pool) -- (head);
  \draw[flow] (head) -- (out);
  \node[io] (noise) at (\cx-55,29) {$p,\lambda$};
  \draw[flow] (noise) -- (head);
\end{tikzpicture}%
}
    \end{picture}}
    \caption{Decoding the surface code with a CNN decoder. (a) The CNN architecture takes measurements with the dual-rail embedding as the input. It then does a 3D convolution of detectors in accordance with their space-time coordinates and produces its output by pooling over all data qubits. Noise parameters $p$ and $\lambda$ are supplied at both the input and the output sections of the decoder. (b) Block logical error rate $p_L$ compared to noise rate $p$ at fixed loss fraction $\lambda=0.98$. The performance of the CNN decoder (dark, circles) is shown in comparison to envelope MLE decoding (light, squares) and envelope matching (light, triangles). The CNN decoder has a nearly identical logical error rate in the measurable regime. (c)--(d) Fitted threshold $p_{\text{th}}$ and distance coefficient $\alpha$ from the ansatz \eqref{eq:pL-ansatz} for multiple loss fractions $\lambda$. }
    \label{fig:surface-code}
\end{figure*}
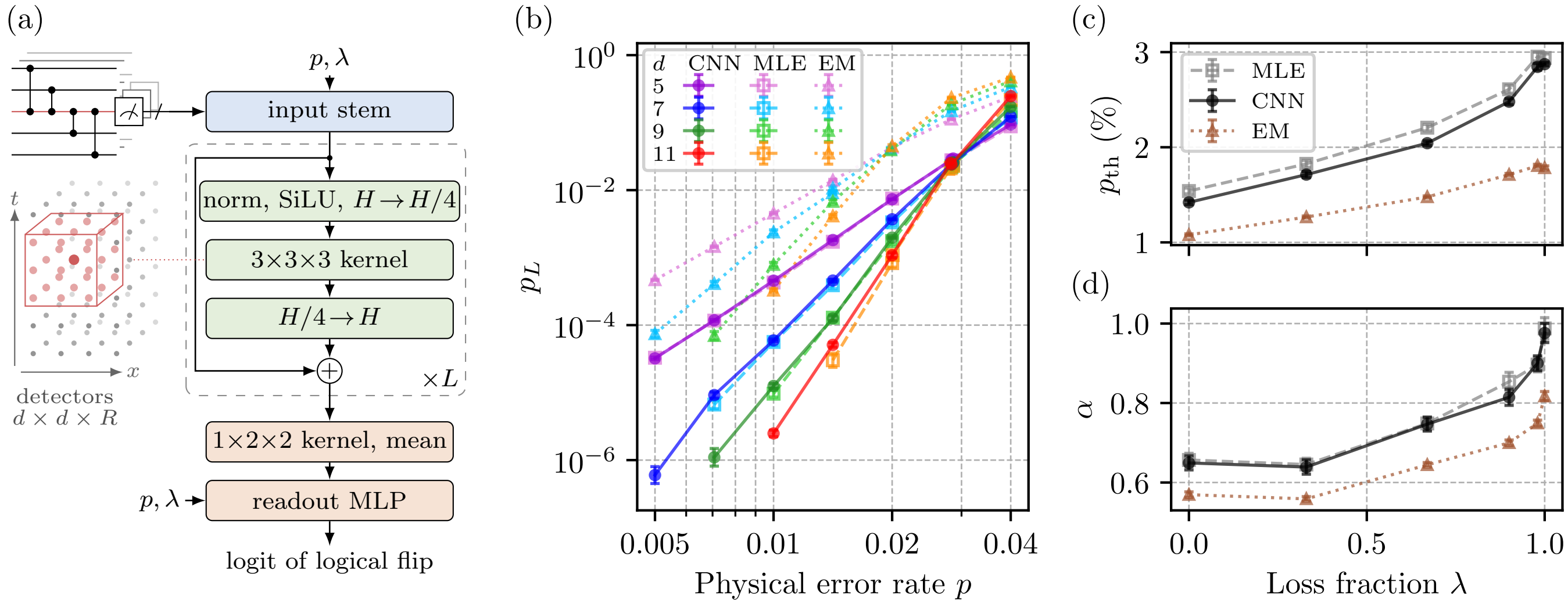

Using \texttt{slim}, we train convolutional neural networks (CNNs) to predict the logical outcome for a surface code with qubit loss in the memory context. Our decoders build on those of Ref.~\cite{gu_scalable_2026}, from which we take the residual convolutional backbone and its extension to Tanner graphs. We use the moonwalking syndrome extraction circuit, specified in Ref.~\cite{pavlovich_erasure_2026}, built with $CZ$ and $H$ gates. 

We use a hybridized noise model with both Pauli and qubit loss errors, parameterized by noise strength $p$ and loss fraction $\lambda$, and include noiseless loss heralding with three-outcome measurements. All errors are connected to a $CZ$ gate, motivated as a model of Rydberg gates for neutral atom qubits~\cite{wu_erasure_2022}. For a given gate, loss errors occur symmetrically on each involved qubit and include a probability $0.07\lambda p$ of occurring on both qubits, reflecting a fixed amount of correlated qubit loss~\cite{yu_locating_2025, perrin_correlated_2026, murri_maximally-matched_2026}. If loss occurs on a single qubit, we apply a $Z$ error on the other qubit with probability $1/2$~\cite{sahay_high-threshold_2023, perrin_quantum_2024, yu_processing_2025, baranes_leveraging_2025}. We additionally apply two-qubit depolarization with strength $(1-\lambda)p$. Additional detail is listed in App.~\ref{app:noise-model}.

We illustrate our architecture in Fig.~\ref{fig:surface-code}a as consisting of three distinct stages. First, we embed the sampled measurement values into the network via an input stem. Each element of the syndrome $\sigma_i$ can be represented as the parity of measurements, which is deterministically $0$ in the absence of noise. The input stem has tied weights for each element of the syndrome and acts on each of its constituent measurements, encoded in the two-bit format
\begin{equation}
    \begin{gathered}
        \ket{0} \to (1, 0)\\
        \ket{1} \to (0, 1)\\
        \ket{L} \to (0, 0),
    \end{gathered}
\end{equation}
where $\ket{L}$ signifies the result from measuring a qubit outside of the computational subspace. An affine map shared by all measurements maps each two-bit pair to two real values $x$, which are each passed through $-\tanh(x/2)$. The measurement products are combined with explicit syndrome parity with a multilayer perceptron of width $32$. The resulting features are embedded into a large dimension $H$ through a trained affine map. An additional trained affine map incorporates the noise parameters $p$ and $\lambda$ through an additive embedding, yielding an ``enriched'' representation of the detector used as input to the convolutional backbone.

The second stage of the architecture consists of $L$ layers of convolution. This uses a distinct $3 \times 3 \times 3$ kernel for each layer, which updates the value of each enriched detector using only nearest neighbors in the spacetime graph. Each layer is a residual bottleneck: a projection from $H$ channels down to $H/4$, a $3\times3\times3$ convolution at this width, and a projection back up to $H$ channels. Each convolution is preceded by a normalization and a SiLU activation, and the result is added to the layer's input with weight $1/\sqrt{2L}$.

The final stage of the architecture is readout. A $1\times2\times2$ convolution maps the four stabilizers around each data qubit to a vector at that qubit. We average these vectors over the data qubits at all space-time coordinates. A perceptron with one hidden layer of width $2H$ and a SiLU activation takes this average, together with the noise parameters, and outputs the logit of the probability that the logical observable flipped. We train on the binary cross-entropy of this logit against the sampled logical outcome.

For our demonstration on the surface code, we train distinct models for each distance $d \in \{ 5, 7, 9, 11\}$. We use $H=256$ for all models ($1.8$--$3.6\times10^{6}$ parameters for $d=5$--$11$), although we believe that increasing $H$ may improve performance at the higher distances (see App.~\ref{app:width-dependence}). We use $L = 2d$ for all models. Since each model is used for a broad range of values for $(p, \lambda)$, both of these noise parameters are supplied to the readout head. In Fig.~\ref{fig:surface-code}b, we show that the models reach logical error rates comparable to those of the envelope MLE baseline~\cite{liu_pauli_2026} for fixed loss fraction $\lambda=0.98$ across a broad range of $p$-values. We additionally show envelope matching~\cite{liu_pauli_2026} as an efficient algorithmic decoder for comparison. Both algorithmic decoders are described in App.~\ref{app:baseline}.

In Figs.~\ref{fig:surface-code}c and \ref{fig:surface-code}d, we show that the surface code models have similar performance to envelope MLE across the full range $0 \leq \lambda \leq 1$. These panels report fit parameters from the ansatz
\begin{equation}
    \label{eq:pL-ansatz}
    p_L \sim \left( \frac{p}{p_{\text{th}}} \right)^{\alpha d + \beta }
\end{equation}
for each $\lambda$. Only data with $p_L < 1.2 \times 10^{-2}$ are used in the fit, in order to capture the subthreshold behavior of interest. The offset $\beta$ is shared between the CNN and envelope MLE fits at each $\lambda$, and the fit for envelope matching is constrained to the same value. Since the same distance-specialized models are used at each loss fraction, each model uses the explicit inputs $p$ and $\lambda$ in place of algorithmically defined priors.

\section{Decoding the Gross Code with Loss}

\begin{figure}[t]
    \centering
    \includegraphics[width=\columnwidth]{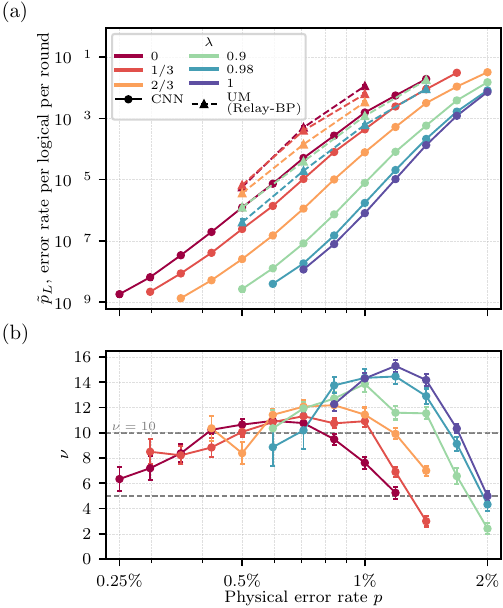}
    \caption{Decoding the Gross code under qubit loss. (a) Logical error rate per logical qubit per round $\tilde{p}_L$ of the CNN (solid, circles) and of the Uniform Mixture decoder (built on Relay-BP) (dashed, triangles) against $p$, for several loss fractions $\lambda$. Both are computed from the block error rate $p_L$, the probability that any of the $k=12$ logical qubits fails over $R=12$ rounds. $1\sigma$ error bars are included and often smaller than markers. (b) Local noise scaling exponent $\nu$ \eqref{eq:nu} of the CNN at each value of $p$, estimated from fault counting for each collected data point. Horizontal guiding lines are shown for both full- and half-distance scaling.}
    \label{fig:gross-code}
\end{figure}

We can generalize the CNN decoder for the surface code to any code with translational symmetry. To exemplify this, we use these methods to decode the $[[144, 12, 12]]$ bivariate bicycle code~\cite{bravyi_high-threshold_2024}, also known as the Gross code, with qubit loss noise. Bivariate bicycle codes have also been studied under erasure-biased noise with gate-resolved erasure information~\cite{pecorari_quantum_2025} and scheduled erasure checks~\cite{bhave_bibieq_2026}. These settings differ from the delayed loss heralding and persistent skipped gates considered here.

In order to halt the persistent effects of loss errors, we use a circuit realization of the Gross code where qubits are periodically teleported. To accomplish this, we implement the ``Walking Gross Code'' which is an application of the techniques specified in Refs.~\cite{mcewen_relaxing_2023, shaw_lowering_2025, shaw_optimising_2026} and adapts the standard syndrome extraction circuit from Ref.~\cite{bravyi_high-threshold_2024}. Explicitly, we reverse the final CNOT on each qubit. These CNOTs pair every $X$-check ($Z$-check) qubit with a distinct data qubit in the left (right) block of the code. The reversal causes the two qubits in each pair to swap data and ancilla roles every cycle. Since the standard circuit uses $144$ ancilla qubits, the walking implementation requires no additional qubits. Each qubit is then measured and reset every other cycle, after at most twelve entangling gates. This circuit has a Pauli distance ($\lambda=0$) of 10 and a loss distance of $\leq 12$. 

We additionally adapt the convolutional portion of the network to respect the structure of the code, following Ref.~\cite{gu_scalable_2026}. This requires that we replace the $3\times3\times3$ space-time convolution with a convolution over the Tanner graph. Each such convolution is built from two linear maps applied in sequence: the first passes the value of each check to the six data qubits in its support, and the second passes the value of each data qubit to the six checks that act on it. Both maps also couple adjacent rounds, and their weights are shared across all translations of the code. Together with a learned self-interaction term, this updates each detector using its own value and those of every detector which it neighbors on the Tanner graph. For our circuit, each update draws on $22$ detectors, the target and its $21$ neighbors, in each of three consecutive rounds. As with the surface code, we apply $L=2d=24$ such layers with width $H=512$ as the bulk of the network ($3.0\times10^{7}$ parameters in total).

Since the Gross code has $k=12$ logical qubits, readout is complicated in comparison to the all-detector pooling used for the surface code. Instead, after convolution, an additional learned layer maps the detector values onto the data qubits using one further step of the Tanner-graph convolution, in which each data qubit is updated from the six checks adjacent to it~\cite{gu_scalable_2026}. We then average over the support of each logical qubit and use an MLP with the same structure as the one used in the surface code to get the output logit. The same perceptron is applied to get each of the $12$ output logits, and we train on the binary cross-entropy averaged over the twelve logical qubits.

We find that the choice of logical representatives affects both training time and eventual performance of the trained network. We get the best performance out of the weight-$12$ representatives defined in Ref.~\cite{bravyi_high-threshold_2024} and measure only the $Z$-memory. Further study is required to assess whether this generating set is optimal and what characteristics are relevant for decoder training.

We show the performance of this network in Fig.~\ref{fig:gross-code}a measured as the logical error rate per logical qubit per round: 
\begin{equation}
     \tilde{p}_L := \frac{1}{2} \left(1-(2(1-p_L)^{1/12}-1)^{1/12}\right)
\end{equation}
where $p_L$ is the block error rate over $R=12$ rounds. We find that the fully heralded case with $\lambda=1$ has an advantage over the Pauli case with $\lambda=0$ of between $10^3$ and $10^4$ for $p \in [7 \times 10^{-3}, 10^{-2}]$. Since heralded loss noise induces a larger logical sensitivity exponent, 
\begin{equation}
    \label{eq:nu}
    \nu := \frac{\partial \left[ \log \tilde{p}_L \right]}{\partial \left[ \log p \right]}
\end{equation}
we expect this advantage to increase in the lower noise regime. 

Our envelope MLE fails to reliably certify a solution in $900$~s, the maximal time we allow. As a reference benchmark, we instead consider the Uniform Mixture (UM) decoder~\cite{murri_maximally-matched_2026}, which uses Relay-BP~\cite{muller_improved_2025} on a matching hypergraph averaged over possible loss configurations. We outperform this benchmark significantly, particularly for large loss fractions. For instance, at $p=0.7\%$ and $\lambda=0.98$, we outperform the UM decoder by a factor $\approx 10^3$.

Notably, the logical error rate of the Gross code under this decoder is not well modeled by a constant power law for each $\lambda$. We instead see that the sensitivity exponent~\eqref{eq:nu} varies with the physical noise rate $p$, consistent with the waterfall regime observed in Ref.~\cite{gu_scalable_2026}. We plot this dependence in Fig.~\ref{fig:gross-code}b using the fault-counting methods developed in Ref.~\cite{fu_efficient_2026} to extract a $p$-local estimate of $\nu$. For each value of $\lambda$, we observe that sensitivity peaks at some high value of $p$ before falling in the low-$p$ regime. Notably, we observe that $\nu > d$ over a range of $p$ when there is a large fraction of heralded loss. We believe this is attributable to typical failures coming from logical representatives of non-minimal weight. 

\section{Discussion}
\label{sec:discussion}

The sampling techniques and decoders presented in this work are tools which expand the scope of studying error correction for loss noise. The ability to study larger codes in a computationally practical way opens the door to more detailed studies of how different implementations of leakage reduction at the hardware level affect the overhead and failure rate at the logical level. This is particularly important in the case of low-density parity-check codes, where deeper syndrome extraction circuits allow loss errors to persist for longer and decrease the specificity of three-outcome measurements. 

The sampling technique demonstrated here is valid for circuits satisfying a ZRC. That is, all zeros is a possible measurement output. This is not satisfied by general Clifford circuits. Whether a broader class of circuits can be efficiently simulated this way is left to future work.

We note that the sampler can be interpreted as a set of measurement results (detectors and loss heralds) calculated from Boolean clauses derived from independent error events (Pauli errors and losses). Decoding is then solving the inverse problem: given the measurement results, find the most likely set of error events. In general, this is a Boolean constraint satisfaction problem. In the absence of losses, the constraints are purely XOR and decoding becomes a graph (or hypergraph) matching problem. While the addition of AND makes the problem more challenging, it is possible that structure identified here may provide a new route to algorithmic decoders with loss.

While in this work we only consider three-outcome loss measurement, it is possible to extend the decoding framework to include mid-circuit erasure detection and imperfect detection. This opens the door to exploring a vast array of hardware-specific optimizations to get the maximal benefit of erasure conversion with minimal experimental overhead.

Another natural direction for further study is investigating methods to verify the optimality of these decoders. Based on our study of the Gross code, power-law scaling is not a generic model of the sub-threshold logical error rate of codes under loss noise. This motivates the development of more sophisticated analytical and heuristic models, which can be used to assess the performance saturation of learned decoders. 

\textit{Note added}--- While finalizing this manuscript, we became aware of related approaches to loss sampling in Refs.~\cite{gidney_css_2026, hou_ellis_2026}.

\textit{Acknowledgments}--- Large language models, including various generations of ChatGPT and Claude, were used during this work. This work was
supported by the U.S. Department of Energy, Office
of Science, National Quantum Information Science Research Centers, Co-design Center for Quantum Advantage (C2QA) under contract number DE-SC0012704, the Army Research Office (W911NF-24-10358), the National Science Foundation (QLCI grant OMA-2120757) and the Gordon and Betty Moore Foundation (grant DOI 10.37807/gbmf12253). Winston Fu is supported by the A*STAR National Science Scholarship.
\textit{Competing interests}--- J.D.T. is a shareholder in Logiqal, Inc.

\bibliography{staples_bib}

\clearpage
\onecolumngrid
\appendix

\section{Noise Model}
\label{app:noise-model}

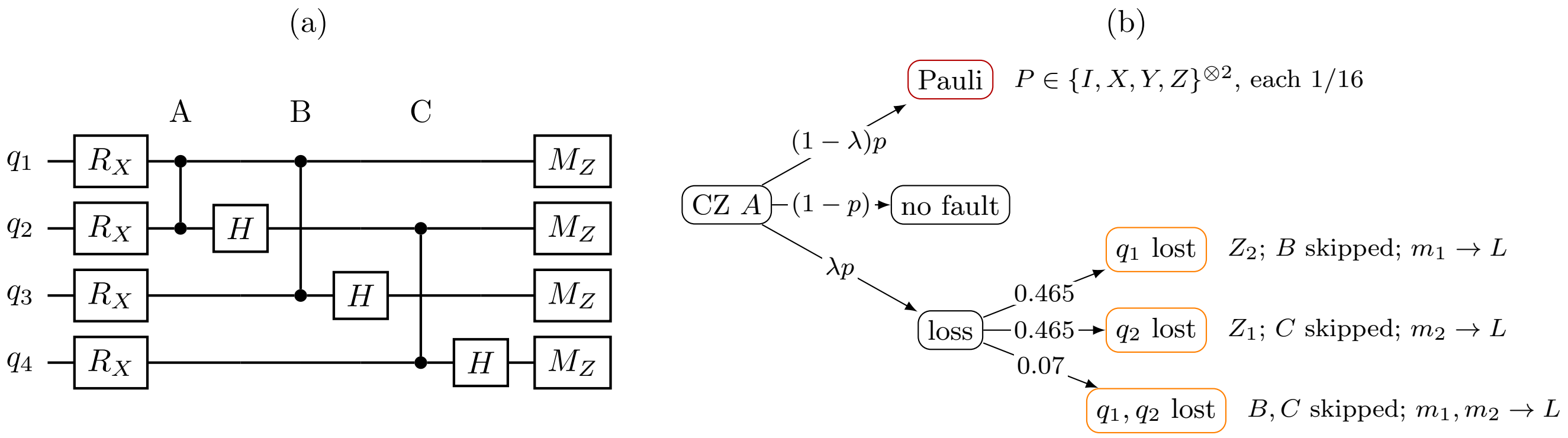
\begin{figure}[ht]
\centering
\begin{minipage}[t]{0.40\textwidth}\centering (a)\end{minipage}\hfill
\begin{minipage}[t]{0.58\textwidth}\centering (b)\end{minipage}\\[4pt]
\begin{minipage}[c]{0.40\textwidth}
\centering
\begin{quantikz}[row sep={0.75cm,between origins}, column sep=0.3cm]
\lstick{$q_1$} & \gate{R_X} & \ctrl{1}\gategroup[1,steps=1,style={draw=none},label style={label position=above, yshift=-0.25cm}]{A} & \qw & \ctrl{2}\gategroup[1,steps=1,style={draw=none},label style={label position=above, yshift=-0.25cm}]{B} & \qw & \qw\gategroup[1,steps=1,style={draw=none},label style={label position=above, yshift=-0.25cm}]{C} & \qw & \gate{M_Z} \\
\lstick{$q_2$} & \gate{R_X} & \control{} & \gate{H} & \qw & \qw & \ctrl{2} & \qw & \gate{M_Z} \\
\lstick{$q_3$} & \gate{R_X} & \qw & \qw & \control{} & \gate{H} & \qw & \qw & \gate{M_Z} \\
\lstick{$q_4$} & \gate{R_X} & \qw & \qw & \qw & \qw & \control{} & \gate{H} & \gate{M_Z}
\end{quantikz}
\end{minipage}\hfill
\begin{minipage}[c]{0.58\textwidth}
\centering
\begin{tikzpicture}[
  every node/.style={font=\small},
  evt/.style={draw, rounded corners, inner sep=3pt, align=center},
  leaf/.style={align=left, font=\footnotesize},
  lab/.style={font=\footnotesize, fill=white, inner sep=1pt},
  level 1/.style={sibling distance=1.4cm, level distance=2.5cm},
  level 2/.style={sibling distance=0.9cm, level distance=2.3cm},
  grow=right, edge from parent/.style={draw, -latex}]
\node[evt] {CZ $A$}
  child {node[evt] (loss) {loss}
    child {node[evt, draw=orange] (both) {$q_1, q_2$ lost}
      edge from parent node[lab] {$0.07$}}
    child {node[evt, draw=orange] (l2) {$q_2$ lost}
      edge from parent node[lab] {$0.465$}}
    child {node[evt, draw=orange] (l1) {$q_1$ lost}
      edge from parent node[lab] {$0.465$}}
    edge from parent node[lab] {$\lambda p$}}
  child {node[evt] (none) {no fault}
    edge from parent node[lab] {$(1-p)$}}
  child {node[evt, draw=red!70!black] (pauli) {Pauli}
    edge from parent[shorten >=3pt] node[lab] {$(1-\lambda)p$}};
\node[leaf, anchor=west, xshift=3pt] at (pauli.east) {$P \in \{I,X,Y,Z\}^{\otimes 2}$, each $1/16$};
\node[leaf, anchor=west, xshift=3pt] at (l1.east) {$Z_2$; $B$ skipped; $m_1 \to L$};
\node[leaf, anchor=west, xshift=3pt] at (l2.east) {$Z_1$; $C$ skipped; $m_2 \to L$};
\node[leaf, anchor=west, xshift=3pt] at (both.east) {$B, C$ skipped; $m_1, m_2 \to L$};
\end{tikzpicture}
\end{minipage}
\caption{(a) Four-qubit circuit with $CZ$ gates $A$, $B$, and $C$. (b) Faults after gate $A$, shown to first order in $p$. The three loss mechanisms and the Pauli mechanism are sampled independently, so multiple faults at the same gate enter at order $p^2$. On the single-loss branches, $Z_1$ and $Z_2$ denote a $Z$ error applied to the surviving qubit with probability $1/2$. A measurement outcome $L$ heralds loss.}
\label{fig:noise-model}
\end{figure}

The noise model used for qubit loss is motivated by the erasure-conversion mechanism of Ref.~\cite{wu_erasure_2022}. We assign all noise to entangling controlled-phase gates. The model is governed by two parameters: the noise level $p$, which linearly scales the probability of all independent error mechanisms, and the loss fraction $\lambda \in [0, 1]$, which sets the split of noise into loss and Pauli noise.

The error mechanisms are described in Fig.~\ref{fig:noise-model}. Consider the $CZ$ gate labeled $A$ in the circuit diagram. We sample four fault mechanisms independently after this gate. With probability $(1-\lambda)p$, we draw a two-qubit Pauli uniformly from $\{I,X,Y,Z\}^{\otimes 2}$. The two single-qubit loss mechanisms each have probability $0.465\lambda p$, and the joint loss mechanism has probability $0.07\lambda p$. A single-qubit loss applies a $Z$ error to the surviving partner with probability $1/2$~\cite{sahay_high-threshold_2023, perrin_quantum_2024, yu_processing_2025, baranes_leveraging_2025}. Concretely, when qubit $1$ is lost, gate $B$ is removed from the circuit and the measurement of qubit $1$ heralds a loss event. Likewise, when qubit $2$ is lost, gate $C$ is removed and the measurement of qubit $2$ heralds the loss.

\section{Zero Reference Condition}
\label{app:trivial-ref-con}

A PFS reports each measurement outcome as a sum of sampled effects from the frame and a reference shot $r$, and is exact for any $r$ that the noiseless circuit returns with nonzero probability~\cite{gidney_stim_2021}. \texttt{slim} uses $r=0$ for every shot, although each shot removes a different set of gates. This is not always valid, since removing circuit elements may result in the all-zero measurement outcome becoming impossible.

In order to use $r=0$ for all shots, it must be true that measuring all zeros is possible for any loss configuration (not counting measurements which indicate loss). For simplicity, we use the stricter sufficient condition below: 

\medskip
\noindent\textbf{Definition --- zero-reference condition (ZRC):} Let $C$ be a circuit comprised of elements $R_X$, $M_X$, $H$ and $CZ$. For a subset $\ell$ of its $CZ$ gates, let $C_\ell$ be $C$ with those gates removed. $C$ satisfies the zero-reference condition if, for every $\ell$, the noiseless circuit $C_\ell$ returns $0$ on every measurement with nonzero probability.
\medskip

In order to determine if a circuit built from these elements satisfies this condition, we convert it into a graph in the following way:

\medskip
\noindent\textbf{Definition --- circuit graph:} For a circuit $C$, cut each wire at each $H$, $M_X$ and $R_X$. The resulting pieces of wire become the vertices of the circuit graph $G(C)$. Two vertices are joined by an edge if there is a circuit element ($H$ or $CZ$) touching both corresponding pieces of wire. We assume that every wire begins with a reset and ends with a measurement.
\medskip

We can use the circuit graph to find the probability of measuring all zeros with the ``sum over paths'' technique used in Refs.~\cite{dawson_quantum_2005, montanaro_quantum_2017, koh_computing_2017}. We can do this by assigning a variable $x_v \in \FF_2$ to each vertex $v \in V_{G(C)}$ in the circuit graph. An assignment to these variables represents a ``classical path'' or, equivalently, a possible trajectory in the computational ($Z$) basis. For an assignment $\vec{x}$ to all variables $x_v$, we can calculate an amplitude using matrix elements for each circuit element:
\begin{equation}
    A(\vec{x}) := \prod_{e \in E_{G(C)}} w_e(\vec{x}),
\end{equation}
where an edge $e$ joining vertices $u$ and $v$ has weight $w_e(\vec{x}) = \bra{x_u x_v} CZ \ket{x_u x_v} = (-1)^{x_u x_v}$ if it comes from a $CZ$, and $w_e(\vec{x}) = \bra{x_v} H \ket{x_u} = (-1)^{x_u x_v}/\sqrt{2}$ if it comes from an $H$. We have also omitted any dependence on the initializations. If we initialized any qubits in the $Z$-basis, only assignments which respected this initialization would have nonzero amplitude. However, since we initialize all $N$ qubits in $\ket{+}$, all amplitudes gain a constant prefactor $(1/\sqrt{2})^N$.

Finding the probability of a measurement outcome in the $Z$-basis is straightforward. We simply sum over all amplitudes corresponding to $\vec{x}$ consistent with the measurement. Let $M$ be the set of vertices with a measurement and let $y_m$ be an observed measurement outcome on some $m \in M$. The probability of observing a set of outcomes $\vec{y}$ is then
\begin{equation}
    \PP_Z(\vec{y}) = 2^{-N} \left| \sum_{\vec{x}} A(\vec{x}) \prod_{m \in M } \delta_{x_m, \ y_m } \right|^2,
\end{equation}
where $\delta_{a, \ b}$ is $1$ if $a = b$ and is otherwise $0$. Measuring in the $X$-basis is slightly more involved. Here, all paths $\vec{x}$ contribute, but the sign of their contribution depends on the measurement. If we measure $\ket{-}$ on $m$, then those paths with $x_m = 1$ inherit a $-1$. The probability of observing the outcomes $\vec{y}$ in the $X$ basis is
\begin{equation} 
    \PP_X(\vec{y}) = 2^{-2N} \left| \sum_{\vec{x}} A(\vec{x}) \prod_{m \in M} (-1)^{x_m \cdot y_m } \right|^2.
\end{equation}
From the above, we can see that the special case of measuring all zeros in the $X$-basis has no signs. Thus, we can only have $\PP_X(0) = 0$ if  $\sum_{\vec{x}} A(\vec{x}) = 0$.  

\medskip
\noindent\textbf{Proposition.} If $G(C)$ is bipartite, then $C$ satisfies the zero-reference condition.

\medskip
\noindent\emph{Proof.} Let $\ell$ be a subset of the $CZ$ gates in $C$, and $C_\ell$ the circuit with them removed. $G(C_\ell)$ is a subgraph of $G(C)$ and so is bipartite. Color the vertices of $G(C_\ell)$ red and blue so that every edge joins two distinct colors. Call the set of red vertices $R$ and the set of blue vertices $B$. We will now make an assignment $\vec{x}_R$ to $R$ and calculate the sum of amplitudes conditioned on this assignment. We can do this easily because, for a fixed $\vec{x}_R$, the amplitude factors over $B$:
\begin{equation}
    \sum_{\vec{x}_B} A(\vec{x}_R, \vec{x}_B) \propto \prod_{b \in B} \left(f_b(\vec{x}_R, 0) + f_b(\vec{x}_R, 1) \right),
\end{equation}
where $f_b$ gives the product of contributions from edges adjacent to some blue vertex $b$. Notably, $f_b$ depends only on the red assignment $\vec{x}_R$ and $x_b$.  

Still in the context of a fixed red assignment, take some $b \in B$. Let $c_b$ be the number of edges joining $b$ to a red vertex with assignment $1$. We then have that $f_b \propto (-1)^{c_b \cdot x_b}$, with some positive prefactor independent of $x_b$. The binomial corresponding to $b$ is then positive if $c_b$ is even and otherwise $0$. So, the sum of amplitudes conditioned on $\vec{x}_R$ is positive if $c_b$ is even for all $b \in B$, and otherwise $0$. Since no red assignment can result in a negative amplitude and $\vec{x}_R = 0$ results in a positive amplitude, we must have $\sum_{\vec{x}} A(\vec{x}) \neq 0$.\hfill $\square$
\medskip

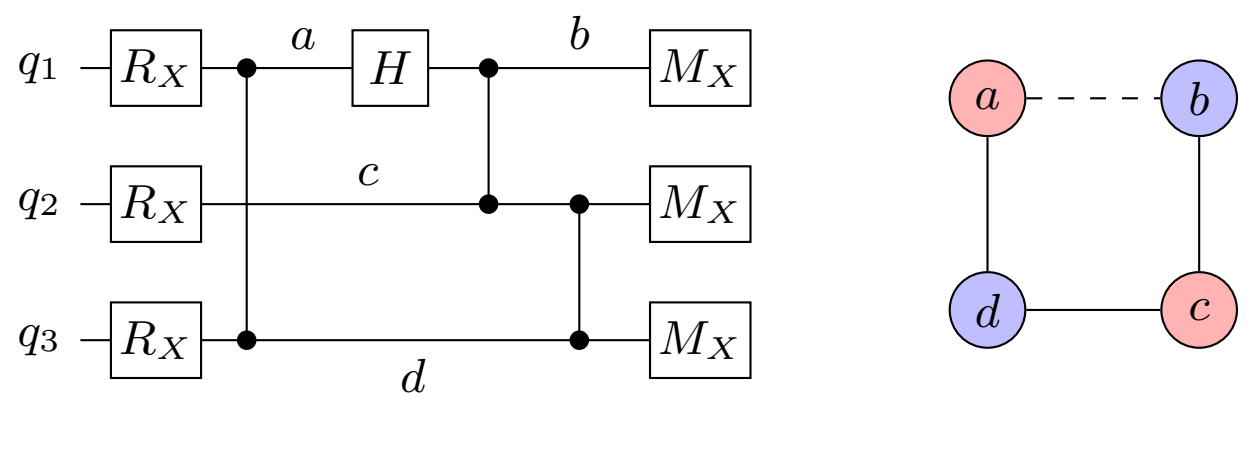
\begin{figure}[ht]
\centering
\begin{tikzpicture}[font=\small, gate/.style={draw, fill=white, minimum size=0.5cm, inner sep=1.5pt}, dot/.style={circle, fill, inner sep=1.3pt}, seg/.style={circle, draw, minimum size=0.5cm, inner sep=0pt}]
  \draw (0,4.0) -- (4.1,4.0);
  \draw (0,3.1) -- (4.1,3.1);
  \draw (0,2.2) -- (4.1,2.2);
  \node[left] at (0,4.0) {$q_1$};
  \node[left] at (0,3.1) {$q_2$};
  \node[left] at (0,2.2) {$q_3$};
  \node[gate] at (0.5,4.0) {$R_X$};
  \node[gate] at (0.5,3.1) {$R_X$};
  \node[gate] at (0.5,2.2) {$R_X$};
  \draw (1.1,4.0) node[dot] {} -- (1.1,2.2) node[dot] {};
  \node[gate] at (2.05,4.0) {$H$};
  \draw (2.7,4.0) node[dot] {} -- (2.7,3.1) node[dot] {};
  \draw (3.3,3.1) node[dot] {} -- (3.3,2.2) node[dot] {};
  \node[gate] at (4.1,4.0) {$M_X$};
  \node[gate] at (4.1,3.1) {$M_X$};
  \node[gate] at (4.1,2.2) {$M_X$};
  \node[above] at (1.5,4.0) {$a$\,};
  \node[above] at (3.3,4.0) {$b$};
  \node[above] at (1.9,3.1) {$c$};
  \node[below] at (2.2,2.2) {$d$};
  \node[seg, fill=red!30] (a1) at (6.0,3.8) {$a$};
  \node[seg, fill=blue!25] (b1) at (7.4,3.8) {$b$};
  \node[seg, fill=red!30] (c1) at (7.4,2.4) {$c$};
  \node[seg, fill=blue!25] (d1) at (6.0,2.4) {$d$};
  \draw[dashed] (a1) -- (b1);
  \draw (b1) -- (c1);
  \draw (c1) -- (d1);
  \draw (d1) -- (a1);
\end{tikzpicture}
\caption{A circuit (left) and its circuit graph (right). The $H$ cuts its wire into the pieces $a$ and $b$ and is shown with a dashed edge; each $CZ$ is shown with a solid edge. The graph is bipartite (red and blue vertices), so the circuit satisfies the zero-reference condition.}
\label{fig:reference-counterexample}
\end{figure}

Since any circuit constructed from gates $\{R_X, R_Z, M_X, M_Z, H, CZ, CX\}$ can be easily converted into a circuit containing only $\{R_X, M_X, H, CZ\}$ by adding $H$ elements, this identification strategy can be immediately generalized to include the larger gate set. Since bipartite graphs can be easily identified, this strategy takes time linear in the number of circuit elements.

We provide a visual example of this method in Fig.~\ref{fig:reference-counterexample}. To the right of the three-qubit example circuit is the corresponding circuit graph, which is colored to demonstrate that it is bipartite. However, if the $H$ gate is removed, the resulting circuit fails the ZRC. In the graph representation, removing $H$ merges vertices $a$ and $b$. As a result, the graph is no longer bipartite.

\section{Algorithmic Decoders}
\label{app:baseline}
We use envelope MLE \cite{liu_pauli_2026} as the baseline for a state-of-the-art, loss-aware algorithmic decoder for the surface code. For each loss event, the decoder constructs a set of local Pauli fault configurations known as the Pauli envelope to represent the effects of qubit loss. For single-qubit loss at a CZ gate, we include a $Z$ fault on the surviving partner with probability $1/2$ in the envelope, as specified by our noise model \cite{sahay_high-threshold_2023, perrin_quantum_2024, yu_processing_2025, baranes_leveraging_2025}. For each shot, different Pauli envelopes are included in the decoding problem based on candidate loss patterns that can explain the measured loss heralds. We weight each candidate pattern by its negative log-likelihood and adjust the objective to account for correlated qubit loss. When both qubits are lost at a CZ gate, one pattern can explain both heralds at a single cost. We solve the resulting integer linear program using Gurobi \cite{gurobi_2025} with a time limit of 60 seconds per shot.

The second algorithmic baseline we test for the surface code is envelope matching \cite{liu_pauli_2026}, which is an efficient approximation of envelope MLE. Starting from the same Pauli envelope construction, we instead use the envelopes to modify and reweight the edges of the loss-free matching graph. We then decode the resulting matching graph using PyMatching \cite{higgott_sparse_2025}.

Finally, for the BB code family, we use the uniform mixture decoder \cite{murri_maximally-matched_2026}. For each heralded loss, we construct a DEM for each candidate loss location and average the DEMs uniformly. The averaged DEM is then combined with the residual Pauli background and decoded using Relay-BP. Envelope MLE was attempted, but with a time limit of $900$~s per shot, time-outs exceeded the rate of logical errors across the regime of interest. 

\section{Training Details}
\subsection{Moonwalking Surface Code}
We train a separate CNN for each code distance with $d+1$ rounds of syndrome extraction. Each optimizer update uses 2,048 shots, minimizing binary cross-entropy between the predicted logical flip probability and the observed logical label. We use Muon~\cite{jordan_muon_2024} for the convolutional backbone weight matrices and Lion~\cite{chen_symbolic_2023} for the remaining parameters, with peak learning rates of $10^{-3}$ and $7\times10^{-5}$ respectively. The learning rates warm up linearly over 100 updates, remain at their peak until update 240,000, and decay linearly to zero by update 480,000.

We vary the physical error rate $p$ and loss fraction $\lambda$ during training. At each training step, we either draw $\lambda$ uniformly from $[0,1]$ or draw $1-\lambda$ log-uniformly from $[10^{-3},1]$. Conditional on $\lambda$, we draw $p$ log-uniformly over a band scaled by the estimated threshold error rate $\tilde{p}_{\text{th}}(\lambda)$ at that loss fraction. We estimate $\tilde{p}_{\text{th}}(\lambda)$ using the envelope MLE decoder. The band is $0.25\tilde{p}_{\text{th}}(\lambda) \leq p \leq \tilde{p}_{\text{th}}(\lambda)$. We multiply the band by $0.25$ for the first 2,400 updates, then increase this multiplier linearly to 1 by update 9,600. The convolution-width sweep uses the same training schedule, with the noise distribution held fixed across widths at each distance.

\subsection{Gross Code}

All Gross code CNN results come from a single model conditioned on $(p,\lambda)$. We minimize binary cross-entropy averaged over all logical outputs, training for 720,000 updates with 2,048 shots per update. We draw $\lambda$ from the same distribution as for the surface code and draw $p$ log-uniformly from $[p_{\max}(\lambda)/4,p_{\max}(\lambda)]$, with the same initial ramp in the band multiplier. At loss fractions
$0$, $0.33$, $0.67$, $0.90$, $0.98$, and $1$, the upper ends of the
band are $0.9000\%$, $0.9751\%$, $1.2355\%$, $1.7048\%$, $2.0000\%$,
and $2.0000\%$, respectively. Between these points we interpolate $1/p_{\max}$ linearly in $\lambda$. The Muon and Lion learning rates warm up over 100 updates to $3\times10^{-3}$ and $2\times10^{-4}$, respectively, remain at their peak until update 240,000, and decay linearly to zero by update 720,000. We use a weight decay of $0.003$.

\clearpage
\section{Latency Analysis}
\label{app:latency}

\begin{figure}[!htb]
    \centering
    \includegraphics[width=6in]{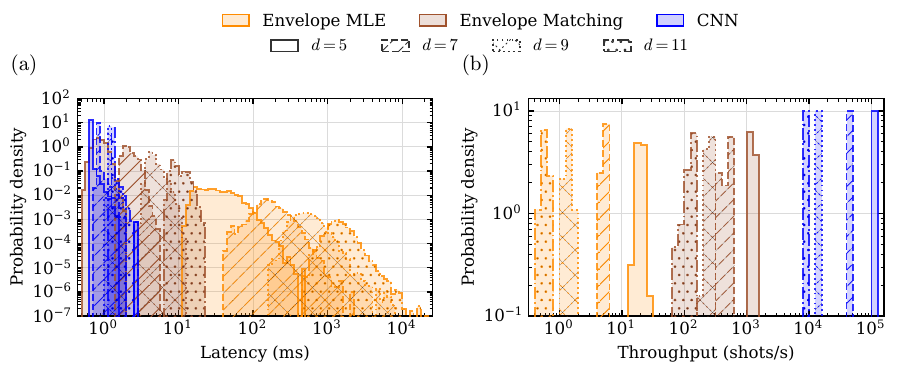}
    \caption{Decoder timing for the moonwalking surface code at $p=2\%$ and $\lambda=0.98$. The $H=256$ CNN ran on an NVIDIA H200 GPU, and the envelope decoders ran on AMD EPYC 9654 CPU cores. (a) Per-shot decoding latency at batch size 1. All three decoders were timed on the same 4,096 shots per distance. (b) Sustained throughput of each decoder, calculated by timing batches of 256 shots.}
    \label{fig:decoder-latency}
\end{figure}

We compared the latency and throughput of the CNN, envelope MLE, and envelope matching decoders. For latency, we measured the wall-clock time to decode one shot of a distance-$d$ moonwalking surface code with $d+1$ rounds (Fig.~\ref{fig:decoder-latency}(a)). Depending on the code distance, the CNN had roughly two to three orders of magnitude lower median latency than envelope MLE.

For sustained throughput, we timed batches of 256 shots for the CNN and envelope decoders (Fig.~\ref{fig:decoder-latency}(b)). The CNN on one H200 sustained roughly four orders of magnitude higher throughput than envelope MLE on one EPYC 9654 core. The CNN also had lower median latency and higher sustained throughput than envelope matching at all four distances.

\section{Dependence on Convolution Width}
\label{app:width-dependence}

\begin{figure}
    \centering
    \includegraphics[scale=1]{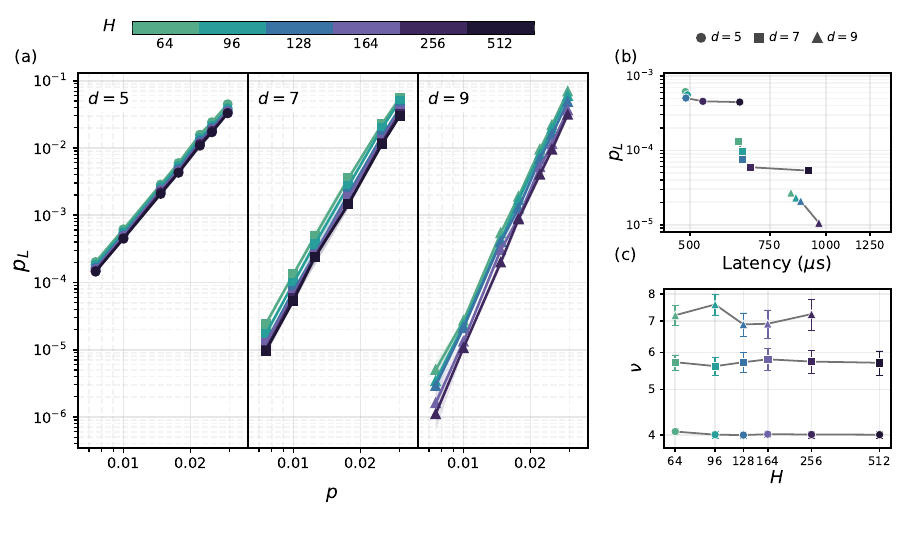}
    \caption{Separate CNNs were trained at $d=5, 7, 9$ with $H = 64, 96, 128, 164, 256$, as well as $H=512$ for $d=5, 7$. All models were trained with $L = 2d$. (a) Logical error rates of each CNN at various values of $p$ between $0.75\%$ and 3\% at $\lambda=0.98$. Each line corresponds to one model. (b) The effect of $H$ on both $p_L$ and latency on NVIDIA H200 GPUs, at different distances at $p=1\%$. Models with $H=164$ are omitted because of their unusually high latency, caused by GPU-specific incompatibilities. (c) The local exponent $\nu$ at $p=1\%$. $\nu$ stays relatively constant with $H$.}
    \label{fig:conv_width}
\end{figure}

We investigated the effects of convolution width $H$ on $p_L$ and inference latency. At $d=5, 7, 9$, we separately trained CNNs with $H = 64, 96, 128, 164, 256$, as well as $H=512$ for $d=5, 7$. All models used $L = 2d$. Fig.~\ref{fig:conv_width}a shows the measured logical error rates between $p=0.75\%$ and 3\%. The results show a consistent reduction in $p_L$ with increasing $H$ over the range of $p$. However, Fig.~\ref{fig:conv_width}b shows that this effect saturates at different $H$ values depending on code distance. The CNNs display distance-dependent saturation behavior. At $d=5$, increasing $H$ from 128 to 256 lowers $p_L$ by 9.7\% but increases latency by 9.0\%. $d=9$ shows a more favorable trade-off, as $p_L$ is lowered by 49\% and latency is raised by 10\% for the same change in $H$. Hence, decoders for higher code distances benefit from a larger convolution width. This motivates our choice of $H=256$ for the surface code. Fig.~\ref{fig:conv_width}c shows that the $p$-local estimate of $\nu$ at $p=1\%$ stays relatively constant with $H$.

\end{document}